\documentclass{pasa}%
\usepackage{graphicx}
\usepackage{lscape}
\usepackage{scalefnt}
\usepackage[]{aalongtable}
\usepackage{scalefnt}

\usepackage{graphicx}

\usepackage[usenames, dvipsnames]{color}

\definecolor{doubt}{RGB}{200, 0, 0}      % dark red
\definecolor{changed}{RGB}{200, 0, 0}      % dark red
\definecolor{todo}{RGB}{212, 132, 46}    % kindda orange

\usepackage[normalem]{ulem}

\usepackage{courier}

\usepackage{mathtools}

\usepackage[mathcal]{euscript}
\newcommand\lstrength{\mathcal{S}}

\usepackage{mathrsfs}
\newcommand\lskovacs{\mathscr{S}}  % "lskovacs" reads "line strength Kovacs"

\title[]{Calculation of molecular line intensity in stellar atmospheres}
\author[Barbuy et al.]{B. Barbuy$^1$, J. Trevisan$^{1}$, A. de Almeida$^1$\\
\affil{$^1$Universidade de S\~ao Paulo, IAG, Rua do Mat\~ao 1226,
Cidade Universit\'aria, S\~ao Paulo 05508-090, Brazil}}%
\jid{PASA}
\doi{10.1017/pas.\the\year.xxx}
\jyear{\the\year}

\begin{document}%
% \input{psfig}

% Next 5 lines define \simless and \simgreat: "less than or approximately
% equal to" and "greater than or approximately equal to".
\newbox\grsign \setbox\grsign=\hbox{$>$} \newdimen\grdimen \grdimen=\ht\grsign
\newbox\simlessbox \newbox\simgreatbox
\setbox\simgreatbox=\hbox{\raise.5ex\hbox{$>$}\llap
     {\lower.5ex\hbox{$\sim$}}}\ht1=\grdimen\dp1=0pt
\setbox\simlessbox=\hbox{\raise.5ex\hbox{$<$}\llap
     {\lower.5ex\hbox{$\sim$}}}\ht2=\grdimen\dp2=0pt
\def\simgreat{\mathrel{\copy\simgreatbox}}
\def\simless{\mathrel{\copy\simlessbox}}
% Next lines define "approximately proportional to"
\newbox\simppropto
\setbox\simppropto=\hbox{\raise.5ex\hbox{$\sim$}\llap
     {\lower.5ex\hbox{$\propto$}}}\ht2=\grdimen\dp2=0pt
\def\simpropto{\mathrel{\copy\simppropto}}

\begin{abstract}
Molecular line intensity calculations are not a straightforward task.
We present a description of the basics for including molecular lines
in synthetic spectra, and of the input data needed. 
%A new version of the spectrum synthesis code is presented.
We aim both at describing ways in which molecular lines are computed in the
 context of photospheres of F-G-K stars,
and to present a new online available version of our code
for spectrum synthesis of cool stars.
 We apply calculations to molecular bands in the ultraviolet,
visible and near-infrared main features, and comparisons with
spectra of reference stars are shown. 
We provide user-friendly tools for the free use of the code PFANT.
The code is available at http://trevisanj.github.io/PFANT.
\end{abstract}
\maketitle%
\section{Introduction}

Molecular bands in the spectra of cool stars need
to be considered in the calculation of synthetic spectra of cool stars,
 first of all because they are present all over the spectra,
 affecting
considerably the continuum, and as well as because there are several
strong  lines. Bands involving C, N, O in particular, are key
indicators of many
important aspects of both stellar evolution and chemical evolution. For example,
photometric filters placed to measure CH, NH, CN, OH, as used in Piotto
 et al. (2015),
can reveal multiple stellar populations in globular clusters.

The main nitrogen indicator used in the literature are CN bands,
 among
which the strongest ones are in the near-UV, in particular CN(0,0) at
388.3\,nm. For the use of CN, one has to previously derive C, available
in the optical from the
CH G-bandhead at 431\,nm or a weak C$_2$(0,0) bandhead at
563.5\,nm. In the near-UV there is the unique available strong NH
bandhead at 336\,nm.  The NH band measurement in important samples
would allow a direct measurement of nitrogen (e.g. Pasquini et al. 2008).
 Note on the other hand,
 that discrepancies have been found in the nitrogen abundances
derived from CN and NH bandheads, still unsolved (see Spite et al.
2005).

Oxygen  has  several strong
OH lines occurring in the near-UV, at $\lambda$$\simless$ 330\,nm,
and in the near-infrared in the H band. In metal-poor
turn-off dwarf stars, the UV OH lines are the only measurable
oxygen indicator.
Furthermore,
molecular bands of CH and CN in the visible and near-infrared are the
best way of deriving the isotopic composition of in particular $12$C/$13$C
(e.g. Smith et al. 2002),
which is an indicator of stellar mixing (e.g. Smiljanic  et al. 2009).

Therefore the inclusion of molecular lines in synthetic spectra
calculations of stars
cooler than $\simless$7000 K is of utmost importance. Line lists of molecular
bands come however in different formats, and as well the different codes
use these lists adapted in different ways.
The main difference in using molecular line lists is the use of either
the Einstein $A$ coefficients,
such as given in Goorvitch et al. (1994) for the near-infrared (NIR)
lines of  CO X$^{1}$$\Sigma^{+}$ system,  which are a sum of all ingredients, or when this
is not available, the different constituents that produce the line.
In the present paper we aim to show how to use these different data.

Codes for spectrum synthesis available include about a dozen of them,
among which
SYNTHE (Kur\'ucz 1970, 1993a), MOOG (Sneden 1973),  the
codes BSYN (Edvardsson et al. 1993), and Turbospectrum (Alvarez \& Plez 1998),
from the Uppsala group, 
and SME (Valenti \& Piskunov 1996; Piskunov \& Valenti 2017), 
among a few others.
In the present paper
we  give details on the updated version of the
spectrum synthesis code PFANT,
that is now available online.

In Sect. \ref{seckappamol} we describe the calculation of molecular
lines of diatomic molecules.
 In Sect. \ref{secpfant} we describe the
spectrum synthesis code PFANT and the line lists used.
In Sect. \ref{secsummary} a short summary is given.

\section{Molecular absorption coefficient}
\label{seckappamol}

In the 60s and early 70s a few authors provided the basis for the understanding
and correct calculation of molecular line formation. 
There are some inconsistencies
in normalization matters among these papers,
which made the subject a hot topic at that time.
These include Schadee (1964, 1967, 1975, 1978), Tatum (1967), and
Whiting \& Nicholls (1974). A particularly clear work on normalization
factors was given later by all these authors gathered in
Whiting et al. (1980). In the following, a description on the calculation
of diatomic molecular lines is given, trying to be as straightforward
as possible, and preparing for the
next Section where these equations are included in the code PFANT presented
in this paper.

The opacity of a molecular line as a result of a transition between two rotational
levels is given by:

\begin{equation}
\begin{multlined}
\label{kappamol}
\kappa_{mol} \; = \\
 {\pi e^{2} \over m_{\rm e} c^2} \; \lambda^{2}_{0 J^{'} J^{''}} \; f \; N^{''}_{nvj} \;
 {H(a,v) \over \sqrt{\pi} \; \Delta \lambda_{D}} \;
  (1 - e^{\; {-hc \over \lambda_{0 J^{'} J^{''}} k T}})
\end{multlined}
\end{equation}

where $f$ = molecular oscillator strength,
$\lambda^{2}_{0 J' J''}$ is the line central wavelength,
$J$ = rotational quantum number,
$v$ = vibrational quantum number,
$n$ = electronic quantum number,
%\sout{f = molecular oscillator strength},
$N''_{nvj}$ = population of the lower level, and
$\Delta\lambda_{\rm D}$ is the Doppler width given by
\begin{equation}
\Delta\lambda_{\rm D} = {\lambda_{0}\over {c}} \sqrt{{2kT\over {M}} + v_{t}^2}
\end{equation}

where $v_{t}$ = microturbulence velocity.
The constants correspond to $c$ = light speed (2.997$\times$10$^{8}$m/s),
$m_{\rm e}$ = mass of hydrogen (1.6737236$\times$10$^{-24}$g), $e$ = electron charge (1.6022$\times$10$^{-19}$C).
 Finally,
$H(a,v)$ = Hjerting's function for line wing broadening,
where $v$ = ${\Delta\lambda\over \Delta\lambda_{\rm D}}$ the damping constant
$\Gamma$ = $\lambda^2$/4$\pi$c$\Delta\lambda_{\rm D}$, where $a$ = damping parameter (see e.g. Gray 2005).

\subsection{Line strength and Einstein coefficients}
\label{einstein}

Einstein (1917) introduced the $A$ and $B$ coefficients to describe
spontaneous emission, and induced absorption and emission. The
Einstein $A$ coefficient is defined in terms of spontaneous emission
from level 2 to level 1 by the probability: $dW = A_{21}dt$. In a radiation
field with a radiation density $\rho_{\nu}$ at frequency $\nu$, a transition
from 1 to 2  has probability $dW = B_{12}\rho_{\nu}dt$,
and also induced emission with
 the probability $dW = B_{21}{\rho}dt$. In a system with $N_1$, $N_2$ atoms
in levels 1 and 2,
the total rate of transitions is: $W_{21} = A_{21}N_2$ for spontaneous
emission,  $W_{12} = B_{12}^{\nu}\rho_{\nu}N_1$ for induced absorption,
and
$W_{21} = B_{21}^{\nu}\rho_{\nu}N_2$ for induced emission.
He also showed that $B_{21}^{\nu} = (\pi^2c^3/\-{h}\nu_{21}^3)A_{21}$
and $B_{12}^{\nu} = (g_2/g_1) B_{21}^{\nu}$, with $g$ = degeneracy factor.

The line intensity
resulting from a spontaneous photon emission in a rotational transition,
from levels ($J'$, $J''$) is

\begin{equation}
E_{J'J''} = N_{J'}h\nu_{J'J''}A_{J'J''}
\end{equation}

where $J'$, $J''$ are the rotational quantum numbers of upper and lower levels
of the transition, respectively.

The Einstein probability coefficient for a rotational transition $J'J''$ between
two levels ($n'v'\Lambda'J'p'$, $n''v''\Lambda''J''p''$) will be:

\begin{equation}
\begin{multlined}
A_{J'J''} = {64\pi^4\nu^3\over 3hc^3 d}
 {\sum_{M'M''} \mid {\bf R}_{n'v'\Lambda 'J'p',n''v''\Lambda''J''p''}  \mid^2} = \\
 {64\pi^4\nu^3\over3hc^3 d} \lstrength
\end{multlined}
\end{equation}

where $\lstrength$ = line strength; $\Lambda$ = electronic sublevel of spin;
$p$ = parity (due to dedoubling of rotational states); $M$ = magnetic
quantum number; $d$ = degeneracy; $R_{ij}$ = elements of the matrix of electric
dipole moment = transition moments; ${\mid {\bf R} \mid}^2$ = transition
probability. Note that each
rotational level $J$ has $(2J+1)$ sublevels characterized by a quantum
number $M$. Each of the two states of $\Lambda$ doubling is considered
a distinct rotational state (according to Whiting \& Nichols 1974
and contrariwise to Tatum 1967 and Schadee 1971).

%VER UNIDADES - JORGE USOU UNIDADES COMPATIVEIS COM GOLDMAN

\subsection{Transition Moments {\bf R}}

The transition moment R is defined as:

\begin{equation}
R_{ij} = \int_{-\infty}^{\infty} \Psi_i^* P  \Psi_j dV
\end{equation}

or, in Dirac's notation,

\begin{equation}
{R_{ij} = < \Psi_{i} \mid {\bf P} \mid \Psi_j >},
\end{equation}

corresponding to transition moments for a transition between two states $(i,j)$
such as $(J',J'')$ or $(v',v'')$ etc., where
$\Psi_i$, $\Psi_j$ are total wave functions of states $i$ and $j$, and
{\bf P} = electric dipole moment.

H\"onl \& London (1925) have shown that the wave function can be separated in
radial and angular parts, such that the transition moment can be written as:

\begin{equation}
\begin{multlined}
\sum_{M'M''} \mid {{\bf R}_{n''v''\Lambda''p''J''}^{n'v'\Lambda'p'J'} \mid^2} =
     \sum_{M'M''} {\mid {\bf R}_{nv\Lambda pJ}\mid^2} =  \\
     \lstrength_{nv\Lambda pJ} = {R^{\rm nv}_{\rm rad}\lskovacs_{\Lambda pJ}}
\end{multlined}
%R^{nv}_{rad}S = R^{ev}_{rad}S
% = {\bf R}_e \mid^2 q_{v'v''}
\end{equation}

The superscript ``nv'' symbolizes the dependence of
$R^{\rm nv}_{\rm rad}$ on the
electronic and vibrational eigenfunctions.

\subsubsection{The Vibronic transition moment R$^{\rm nv}_{\rm rad}$}

The Born-Oppenheimer approximation (1927)
 allows the wave function of a molecule to be broken into two components,
 the electronic and the nuclear (vibrational
and rotational), given that the electronic wave function varies very slowly with
the nuclear coordinates. Therefore

\begin{equation}
\Psi_{\rm total} = \Psi_{\rm electronic}\times\Psi_{\rm nuclear}
\end{equation}

The dipole moment can thus be broken into two components,
electronic and nuclear, i.e., ${\bf P} = {\bf P}_{\rm e} + {\bf P}_{\rm N}$.
The transition moment becomes then

\begin{equation}
{\bf R}_{\rm rad} = \int_{-\infty}^{\infty} \Psi_{\rm e}^{'*}\Psi_{\rm v}^{'*}
({\bf P}_{\rm e} +  {\bf P}_{\rm N})
  \Psi_{\rm e}^{''}\Psi_{\rm v}^{''} dV
\end{equation}

and $dV$ is defined by electronic and nuclear coordinates: $dV =
dV_{\rm e}dV_{\rm N}$ or, since $\Psi_{\rm v}$ and P$_{\rm N}$ are functions only
of the internuclear distance $r$, the formula
can be reduced to a radial dependence only, therefore

\begin{equation}
\begin{multlined}
{\bf R}_{\rm rad} = \int_{-\infty}^{\infty}
\Psi_{\rm v}^{'*} (\Psi_{\rm e}^{'*} {\bf P}_{\rm e} \Psi_{\rm e}^{''} dV_{\rm e})\Psi_{\rm v}^{''} dr + \\
\int_{-\infty}^{\infty} \Psi_{\rm v}^{'*}  {\bf P}_N (\Psi_{\rm e}^{'*}\Psi_{\rm e}^{''*} dV_{\rm e})
\Psi_{\rm v}^{''} dr
\end{multlined}
\end{equation}

The wave functions of different electronic states are orthogonal, therefore
the second term of this equation can be canceled. The electronic transition
moment can be defined as:

\begin{equation}
{\bf R}_{\rm e} = \int_{-\infty}^{\infty} \Psi_{\rm e}^{'*} {\bf P}_{\rm e} \Psi_{\rm e}^{''} dV_{\rm e} =
< \Psi_{\rm e}^{'*} \mid {\bf P}_{\rm e} \mid \Psi_{\rm e}^{''} >
\end{equation}

\begin{equation}
{\bf R}_{\rm rad} = \int_{-\infty}^{\infty} \Psi_{\rm v}^{'*} {\bf R}_{\rm e} \Psi_{\rm v}^{''} dr
\end{equation}
%{\bf R}$_{\rm rad}$ = $\langle$ $\Psi_v^{'*} {\bf P}_v \Psi_v^{''} $ $\rangle$

where
${\bf R}_{\rm e}$ can be measured in laboratory experiments, or computed theoretically.
As a first approximation, we can admit that ${\bf R}_{\rm e}$ = constant, and a mean
value can be adopted:

\begin{equation}
{\bf R}_{\rm rad} = {\bf R}_{\rm e} \int_{-\infty}^{\infty} \Psi_{\rm v}^{'*}  \Psi_{\rm v}^{''} dr
\end{equation}

where the Franck-Condon factor is

\begin{equation}
q_{v'v''} = \left | \int_{-\infty}^{\infty} \Psi_{\rm v}^{'*}  \Psi_{\rm v}^{''} dr \right |^2,
\end{equation}

and the band intensity ({\it force de bande}) is

\begin{equation}
  R_{\rm rad}^{\rm nv} = R_{\rm rad}^{\rm ev} = \mid {\bf R}_{\rm e} \mid^2 q_{v'v''}
\end{equation}

\subsubsection{The line strength $\lstrength$}

The line strength for rotational lines will be (Whiting et al. 1980):

\begin{equation}
\lstrength = \mid {\bf R}_{\rm e} \mid^2 q_{v'v''} \lskovacs_{J'J''}
\end{equation}

with $\lskovacs_{J'J''}$ = $\lskovacs_{\Lambda pJ}$ = H\"onl-London factor (HLF), giving the relative
intensity of the different rotational transitions. Depending on the Hund's coupling
associated with each case, formulae for computing the HLF are tabulated by
Kov\'acs (1969).

The normalization of the HLF is as follows, valid for all cases of
$\Lambda$-S coupling (Whiting et al. 1980):

\begin{equation}
\label{eq:norm}
\sum {\lskovacs _{J'J''}(J)} = (2-\delta_{0,\Lambda'+\Lambda''})(2S+1)(2J+1),
\end{equation}

where the sum is carried out for all branches and fixed $J$.
The Kronecker delta $\delta_{0,\Lambda'+\Lambda''}$ equals 1 for $\Sigma - \Sigma$ transitions and 0 otherwise. The term
$(2-\delta_{0,\Lambda'+\Lambda''})(2S+1)$ represents the degeneracy of the
electronic state,  $(2J+1)$ %\sout{that of each}
 the degeneracy of each rotational level, and
$(2S+1)$ represents the multiplicity of the transition: 1 for singlet,
 2 for doublet, etc.

% \textcolor{changed}
% {Note the distinction between symbols
%     $\lstrength$ (line strength),
%     $\lskovacs$ = ${\lskovacs_{J'J''}$ (HLF as tabulated by Kov\'acs (1969)), and
%     $S$ (total spin angular momentum).}

\subsubsection{The oscillator strength $f$}

The oscillator strength $f$ is defined as:

% ${f_{J'J''} =  f_{n''v''\Lambda''p''J''}^{n'v'\Lambda'p'J'}  = f_{nv\Lambda pJ} =}$

% \begin{equation}
%  {8\pi^2m_e\nu\over {3he^2 d}}
%  {\sum_{M'M''} \mid {\bf R}_{n'v'\lambda'J'p',n''v''\lambda''J''p''}  \mid} =
% {8\pi^2m_e\nu\over {3he^2 d}} {\it S}
% \end{equation}

\begin{equation}
\begin{multlined}
f = f_{J'J''} =  f_{n''v''\Lambda''p''J''}^{n'v'\Lambda'p'J'}  = f_{nv{\Lambda}pJ} = \\
{8\pi^2m_{\rm e}\nu\over {3he^2 d}}
{\sum_{M'M''} \mid {\bf R}_{n'v'\lambda'J'p',n''v''\lambda''J''p''}  \mid}{^{2}} = \\
{8\pi^2m_{\rm e}\nu_{J'J''}\over {3he^2 d}} \mid {\bf R}_{\rm e} \mid^2 q_{v'v''} \lskovacs_{J'J''} =
{8\pi^2m_{\rm e}\nu\over {3he^2 d}} \lstrength
\end{multlined}
\end{equation}

Therefore, by combining equations (4) and (18), we get

\begin{equation}
{A\over f} = {8\pi^2\nu^2e^2\over c^3m_{\rm e}}
\end{equation}

%NESTE A, VER PESOS ESTATISTICOS QUE FORAM CORTADOS - nao foram cortados,
%pois tem o d = degenerescencia

%The oscillator strength $f$ is:

%\begin{equation}
%\begin{multlined}
% f = f_{J'J''} = \\ {8\pi^2m_{\rm e}\nu\over {3he^2 d}} \lstrength =
%{8\pi^2m_{\rm e}\nu_{J'J''}\over {3he^2 d}} \mid {\bf R}_{\rm e} \mid^2 q_{v'v''} \lskovacs_{J'J''}
%\end{multlined}
%\end{equation}

Designating $f_{v'v''}$ = vibronic oscillator strength, we have:

\begin{equation}
f_{v'v''} = {8\pi^2m_{\rm e}\nu\over {3he^2}} R^{\rm nv}_{\rm rad} =
{8\pi^2m_{\rm e}\nu_{J'J''}\over {3he^2}} \mid {\bf R}_{\rm e} \mid^2 q_{v'v''},
\end{equation}

therefore

\begin{equation}
f = {f_{v'v''} \lskovacs_{J'J''} \over d}.
\end{equation}

The degeneracy factor $d$ is $(2J''+1)$ for a single rotational line:

\begin{equation}
f^{n'v'\Sigma'p'J'}_{n''v''\Sigma''p''J''} = {f_{v'v''} \lskovacs_{J'J''} \over (2J''+1)}
\end{equation}

%And total oscillator strength of all transitions from one vibronic
%state (n''v''J'') = degenerate rotational state of vibronic state
%(n'v') ----????

%\begin{equation}
%f_{v'v''} = f^{n'v'}_{n''v''J''} = f_{v'v''}
%\sum_{J'}\sum_{\Sigma'\Sigma''}\sum_{p'p''} {\rm S}_{\Sigma{pJ}}
%\over{(2-\delta_{0,\Lambda''}(2S+1)(2J''+1)}
%\end{equation}

%\begin{equation}
%{\nu_{J'J''}\over{\nu_{v'v''}} f_{v'v''}S_{\Lambda pJ}\over{(2J''+1)}}
%\end{equation}

%with n = electronic state, v = vibration quantum number; S = global spin
%quantum number, where (nv) consists of (2S+1) vibronic degerate levels
%(nv$\Sigma$); each of these with the series of rotational states
%(nv$\Sigma$J), with J = quantum number of the total angular momentum.
%The $\Lambda$-degeneracy of electronic state n, leads to the so-called
%$\Lambda$-dedoubling of each rotational state.

%??? (nv$\Sigma$pJ) = simple rotational state (p = parity, Tatum 1967);
%(nvJ) = general rotational state encompasses all rotational states
%(nv$\Sigma$J) with same n, same v, same J; the degeneracy of
%(nvJ) is (2S+1).

The molecular absorption coefficient can then be expressed by:

\begin{equation}
\begin{multlined}
\kappa_{mol} \; = \; {\pi e^{2} \over m_{\rm e} c^2} \; \lambda^{2}_{0 J^{'} J^{''}} \;
    f_{v'v''} {\lskovacs_{\Lambda{pJ}}\over{(2J''+1)}}  \; N^{''}_{nvj} \; \times \; \\
{H(a,v) \over \sqrt{\pi} \; \Delta \lambda_{D}} \;
    (1 - e^{\; {-hc \over \lambda_{0 J^{'} J^{''}} k T}}),
\end{multlined}
\end{equation}

%where 
%\begin{equation}
%f = {f_{v'v''} \lskovacs_{\Lambda{pJ}}\over{(2J''+1)}}.
%\end{equation}

%\begin{equation}
% {8\pi^2m_e\over {3he^2}} \nu_{J'J''}q_{v'v''}f_{el} {S_{\Lambda pJ}
%\over {(2J''+1)}}
%\end{equation}

%where
%\begin{equation}
% f_el = \sum_{M'M''} \mid \mid R_{e}\mid^2\over {2-\delta_{0,\Lambda ``}} (2S''%+1)
%\end{equation}

\subsection{Calculation of the lower level population  $N^{''}_{nvj}$}

Consider the equilibrium reaction A + B $\rightleftharpoons$ AB,
 between two atoms A and B, of the AB molecule, which is expressed by
the relation:

\begin{equation}\label{eq:nanb0}
\begin{multlined}
{n({\rm A})n({\rm B}) \over n({\rm A})} =
    {(\sum_{i} g_i e^{-\epsilon_i/kT})_{\rm A} (\sum_{i} g_i e^{-\epsilon_i/kT})_{\rm B} \over
(\sum_{i} g_i e^{-\epsilon_i/kT})_{{\rm A}{\rm B}}} = \\
{Q_{{\rm total}_{\rm A}} Q_{{\rm total}_{\rm B}} \over Q_{{\rm total}_{{\rm A}{\rm B}}} }
\end{multlined}
\end{equation}

where
$g_{i}$ = statistical weight, $n({\rm X})$ = number of particles X in a volume V(X),
$Q_{\rm total} = Q_{\rm translational}Q_{\rm internal} = (\sum_{i} g_i e^{-\epsilon_i/kT})$ =
total partition function, and the energies
($\epsilon_i$)$_{\rm A}$, ($\epsilon_i$)$_{\rm B}$ are defined with respect
to the fundamental state, i.e.,
$(\epsilon_i)_{\rm A,B} = [(\epsilon_0)_{\rm fundamental state} + (\epsilon_i)_{\rm exc})]_{\rm A,B}$.
The AB molecule energies $\epsilon_i/kT$$_{\rm AB}$ are given relative to the
vibrational fundamental state (quantum number $v=0$), of the fundamental
electronic state of the molecule:
$(\epsilon_i)_{\rm AB} = [(\epsilon_0)_{\rm fundamental state}(v=0) + (\epsilon_i)_{\rm exc}] _{\rm AB}$.

By rewriting eq. \eqref{eq:nanb0}, we have

\begin{equation}
{n({\rm A})n({\rm B}) \over n({\rm A})} =
    {Q_{\rm A} Q_{\rm B} \over Q_{\rm AB}}
    e^{ -(\epsilon_{0_{\rm A}} + \epsilon_{0_{\rm B}} - \epsilon_{0_{\rm AB}})/kT}
\end{equation}

where $\epsilon_{0_A} + \epsilon_{0_B}  - \epsilon_{0_{AB}} = D_{0}$,
and $D_{0}$  is the dissociation energy of molecule AB.

Noting that  $Q_{\rm translational} = ({2\pi mkT\over h^2})^{3/2}V$, we get:

\begin{equation}
\begin{multlined}
{n({\rm A})n({\rm B}) \over n({\rm A})} =
\left ({2\pi mkT\over h^2}\right )^{3/2}{V_A V_B \over V_{\rm AB}} \mu_{\rm AB}^{3/2} \; \times \\
    {Q_{\rm int_A}Q_{\rm int_B} \over Q_{\rm int_{AB}}} e^{-D_0/kT},
\end{multlined}
\end{equation}

where $\mu_{\rm AB} = m_{\rm A} m_{\rm B}/ m_{\rm AB}$ is the reduced mass.

On the other hand, the definition of the dissociation pressure constant is:

\begin{equation}
{K_{\rm AB}^{P}} \; = \; {p_{\rm A} \; p_{\rm B} \over p_{\rm AB} }
\end{equation}

where $p_{\rm A}$, $p_{\rm B}$, and $p_{\rm AB}$ are the partial pressures of particles A, B, and AB.

\begin{equation}
{p_{\rm A} \; p_{\rm B} \over p_{\rm AB} } \; = \;  \;
\left({2 \pi k T \over h^{2} } \; \mu_{\rm AB} \right)^{3/2} \;
 {Q_{int_{\rm A}} \; Q_{int_{\rm B}} \over Q_{int_{\rm AB}} } \;
 kT \; e^{\; {-D_{0} \over kT}}
\end{equation}

Considering now the Boltzmann equation for the population of the level $nvJ$,
 $(N_{\rm AB})_{nvJ}$ being the number of molecules AB in the electronic level $n$,
 vibrational level $v$, and rotational level $J$, per volume unit, and $N_{\rm AB}$
 the total number of molecules AB per volume unit, we have:

\begin{equation}
{(N_{\rm AB})_{nvJ} \over N_{\rm AB}} = { (Q_{int_{\rm AB}})_{nvJ} \over Q_{int_{\rm AB}}}
\end{equation}

and one obtains:

\begin{equation}
{(N_{\rm AB})_{nvJ} \over N_{\rm AB}} \; = \; \left({2 \pi k T \over h^{2} }
\; \mu_{\rm AB} \right)^{-3/2} \; {N_{A} N_{B} \over Q_{int_{A}} \; Q_{int_{B}}} \;
Q_{int_{\rm AB}} \; e^{\; {D_{0} \over kT}}
\end{equation}

\vspace{15 pt}

\subsection{The partition functions Q{\boldmath$_{int_{\rm AB}}$}}

\vspace{10 pt}

Considering that for a molecule the electronic, vibrational, and rotational energies are
independent of each other, Q$_{\rm int}$ may be written as the product:

\begin{equation}
Q_{\rm int} = Q_{\rm el} \; Q_{\rm vibr} \; Q_{\rm rot}
\end{equation}

where:

\begin{eqnarray}
Q_{\rm el} & = & g_{\rm el} \; e^{\; {-T_{\rm e}hc \over kT} } \\
Q_{\rm vibr} & = & g_{\rm vibr} \; e^{\; {-G(v)hc \over kT}} \\
Q_{\rm rot} &  = & g_{\rm rot} \; e^{\; {-F(J)hc \over kT} }
\end{eqnarray}

and,

(a) g$_{\rm el}$, $g_{\rm vibr}$, and $g_{rot}$ are the statistical weights of the electronic,
vibrational, and rotational levels, respectively. The product of the statistical weight
$g_{\rm el}$, $g_{\rm vibr}$, and $g_{\rm rot}$ are given for each case of molecular transition
by Tatum (1967);

(b) $T_{\rm e}$ = term value of electronic energy of a multiplet, in first aproximation it is given
by
\begin{equation}
T_{\rm e} = T_{0} + A \Lambda \Sigma,
\end{equation}
where
$T_{0}$ = value of the term if the spin is negligible;
$A$ = constant of spin coupling;
$\Lambda$ and $\Sigma$ = respectively,
orbital electronic angular momentum and spin angular
momentum, projected in the internuclear axis (not to confound with other
$A$ and $\Lambda$ defined in Sect. \ref{einstein}).

(c) $F(J)$ = term of rotational energy, given by
\begin{equation}
F(J) = B_{v}J(J+1) - D_{v}J^{2}(J+1)^{2} \ldots,
\end{equation}

where $B_{v}$ = rotational constant, $D_{v}$ = centrifugal distortion constant.

Note: $F(J)$ is given with respect to the lowest level,
i.e., $F(J)$ is in reality $F(J) - F(1)$.

(d) $G(v)$ = term of vibrational energy.

A diatomic molecule, by the vibrational movements of its nuclei, can be treated as an anharmonic
oscillator with a potential $U$ approximately given by:

\begin{equation}
U \; = \; f(r - r_{e} )^{2} \; - \; g(r - r_{e})^{3}
\end{equation}

where: $g \ll f$ and $r_{e}$ is the internuclear distance in equilibrium.

Introducing $U$ in the wave equation, one obtains the following energy eigenvalues:

\begin{equation}
\begin{multlined}
E_{v}=hc\; \{\omega_{e} (v+{1\over2} ) \; - \;
             \omega_{e}x_{e} (v+{1 \over 2} )^{2} \; + \; \\
             \omega_{e}y_{e} (v+{1 \over 2} )^{3} \; + \;
             \ldots \}=hc \; G(v)
\end{multlined}
\end{equation}

The values of the anharmonic constants $\omega_{e}$, $\omega_{e}$x$_{e}$, and $\omega_{e}$y$_{e}$ are
tabulated in the literature (for example, in Huber \& Herzberg 1979, or
in the NIST Chemistry web book at http://webbook.nist.gov/chemistry).

Note: $G(v)$ is given with respect to the zero point energy, i.e., $v$ = 0, so in reality one uses
$G_{0}(v) = G(v) - G(0)$.

A discussion on the multiple forms used for partition functions is
presented in Popovas (2014).

\section{PFANT: calculation of synthetic spectra}
\label{secpfant}

The spectra were computed using the code PFANT
(http://trevisanj.github.io/PFANT). The first
version of this Fortran code, named FANTOM, was developed by Spite
(1967) for the calculation of atomic lines. Barbuy (1982) included
the calculation of molecular lines, implementing the
dissociative equilibrium by Tsuji (1973) and molecular line
computations as described in Cayrel et al. (1991). It has been
further improved for calculations of large wavelength coverage
and inclusion of hydrogen lines as described in Barbuy
et al. (2003), where M.N. Perrin had an important r\^ole,
wherefrom comes the P in PFANT. Coelho et al. (2005) 
unified the code  to contain
 the full line lists from the ultraviolet
(Castilho et al. 1999), together with the visible
(Barbuy et al. 2003), and near-infrared (Mel\'endez et al. 2001).

Given a stellar model atmosphere and lists of atomic
and molecular lines, the code computes a synthetic spectrum
assuming local thermodynamic equilibrium (LTE).

\subsection{New changes to the PFANT code}

During 2015-2017, the code was entirely upgraded with homogeneization
of language to Fortran 2003, optimization for speed, improved
error reporting in case of problems with the input data files, and
command-line configuration options. The code now has a manual including
installation and tutorial sections.

\begin{figure}
\centering
\fbox{
\includegraphics[width=\columnwidth, keepaspectratio]{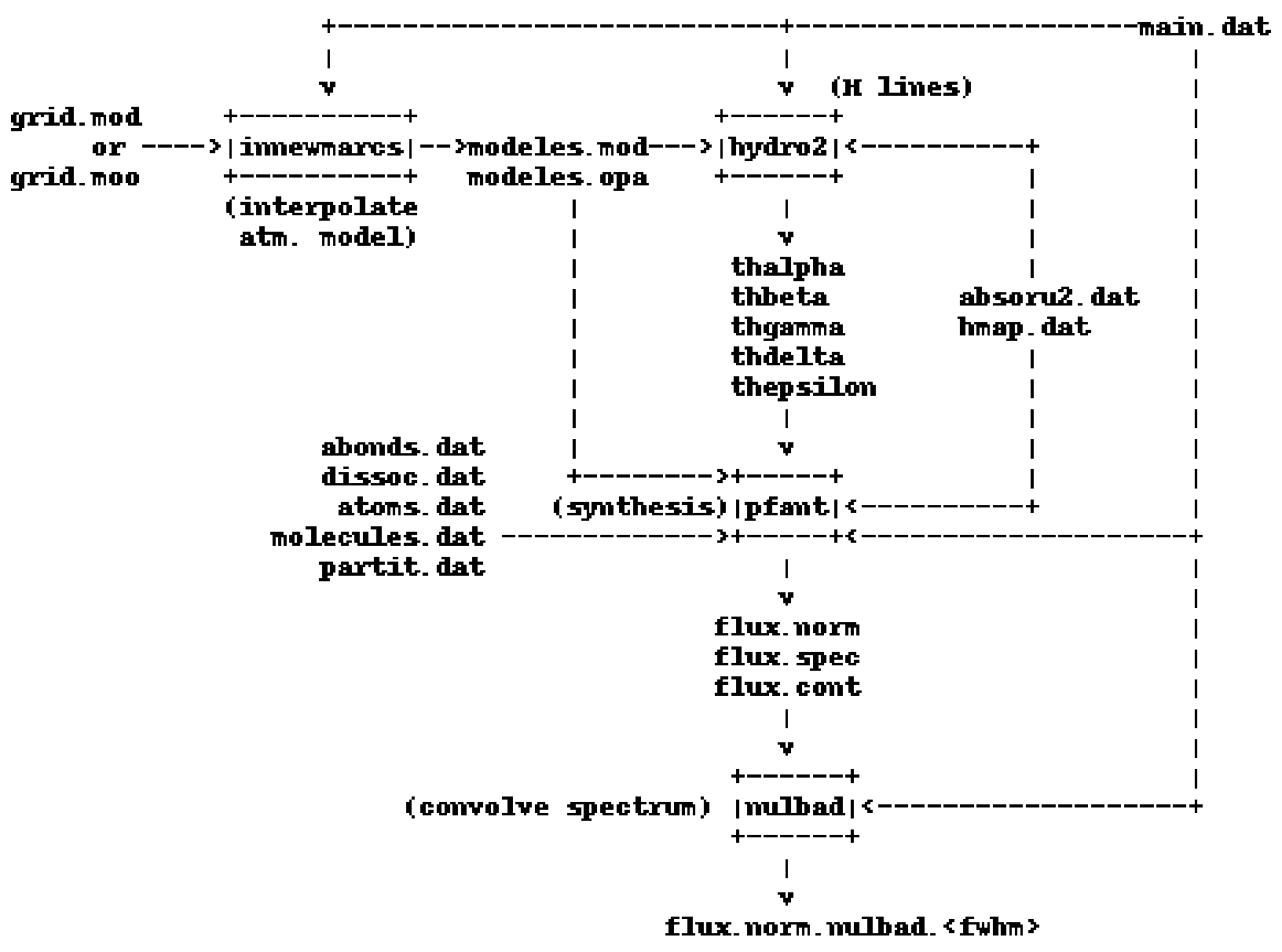}
}
\caption{
PFANT workflow. Boxes represent the Fortran binaries; unboxed text represent input/output files. {\it Source: PFANT manual.}}
\label{workflow}
\end{figure}

Important existing accessory codes in Fortran were also upgraded
and incorporated to form a four-step workflow as follows (Figure \ref{workflow}):

\begin{enumerate}
\item \texttt{innewmarcs} -- interpolate a grid of MARCS atmospheric models
(Gustafsson et al. 2008),
 to generate star-specific model (the original code comes from the Meudon research group);
\item \texttt{hydro2} -- calculate hydrogen lines profiles (Praderie 1967);
\item \texttt{pfant} -- calculate synthetic spectrum;
\item \texttt{nulbad} -- convolve synthetic spectrum with a Gaussian profile to simulate spectral measurement.
\end{enumerate}

The list above describes commands to be typed in a computer terminal.
All commands may be invoked with the \texttt{--help} option.

\subsection{Input and output data}

This section describes the most relevant data files needed or generated
during PFANT code execution. The following subsections refer to these
files by their default names, which can be changed via command-line,
if necessary.

\subsubsection{Stellar parameters and abundances}

Stellar parameters and chemical abundances are defined in files
\texttt{main.dat} and \texttt{abonds.dat}. The code available for download
includes reference solar abundances that can be chosen to be those
from Asplund et al. (2009), Grevesse et al. (1996),
Scott et al. (2015a,b) and Grevesse et al. (2015), or Grevesse et al. plus
A(O)=8.76 from Steffen et al. (2015).

\subsubsection{Model atmospheres}

\texttt{innewmarcs}, the grid interpolation code, allows to call for a
chosen grid among the MARCS
model atmospheric grids, and interpolates for the stellar parameters
needed, creating files named \texttt{modeles.mod} and \texttt{modeles.opa}.
If these parameters, i.e.,  $\mathbf{v}$ = (T$_{\rm eff}$, log~g, [Fe/H])
are outside the grid, the code does not extrapolate, but either stops, or
if forced to run (\texttt{innewmarcs --allow T}), projects $\mathbf{v}$
the onto closest grid ``wall'', then interpolates.

It is possible to replace the grid among the different options given
in MARCS. To create a new grid, it suffices to download the atmospheric
models of interest (files ``.mod'' and ``.opa'')
from the MARCS website (http://marcs.astro.uu.se), place
them in a single folder, and run \texttt{create-grid.py} (a Python script
included with the \texttt{pyfant} package, described in Sect. \ref{secpyfant}).

\subsubsection{Hydrogen lines profiles}

The Balmer, Paschen and Brackett series line data are included in file \texttt{hmap.dat}.
For each hydrogen line, the central wavelength, the levels of the transition,
excitation potential, and a constant related with the oscillator strength
are tabulated.
Hydrogen lines profiles are computed through a revised version of the
code presented in Praderie (1967) (\texttt{hydro2}), generating separate
files for each hydrogen line.

It is well-known that hydrogen lines in red giant stars cannot 
be well-reproduced.
Cayrel et al. (1991) suggested that the bottom of the lines would be due to chromospheric
layers, which are not taken into account in stellar atmosphere models.
 For higher order Balmer lines the overlapping of lines is taken into
 account,
but combined with the Balmer decrement present in the continuum opacities,
 we get a bump in the continuum at 3650-3662 {\rm \AA}. 
This particular region should be used with caution.

\subsubsection{Continuum opacity}

The code \texttt{PFANT} computes the continuum in pure absorption.
It takes into account absorption due to H$^{-}$, H, H$_2^{+}$, He
and He$^{+}$, Rayleigh diffusion by H and H$_2$ and Thomson diffusion
by electrons. A validation of these continuum calculations was
carried out by Trevisan et al. (2011), through comparisons
with the Uppsala code BSYN (Edvardsson et al. 1993, and updates
since then). All
steps of the calculation were carefully compared: optical depths
of lines, continuum opacities $\kappa_{\rm c}$,
line broadening, and final abundances.

%The dominating opacity source is the H − bound-free absorp-
%tion. The two codes consider different calculations for the H −
%photo-detachment cross section σ λ . The ABON2 code (Spite
%1967) adopts Geltman (1962) calculations, represented by the
%polynomial expressions from Gingerich (1964), while cross sec-
%tions from Wishart (1979) are adopted in the BSYN/EQWI code
%(Edvardsson et al. 1993). Using these sources and considering a
%Fe i line at 5861 Å and the solar atmosphere model, we found
%that differences in the continuum absorption are up to 4% in the
%upper atmospheric layers (τ 5 < −1), and less than 1% in the
%bottom of the photosphere (τ 5 > 0). We updated the ABON2
%code (Spite 1967) using new cross sections calculations from
%Johny (1988), which improve the agreement of κ c between these
%two codes to <1% at the to <1% at the upper layers with τ 5 < −1 and ∼2% at
%τ 5 > 0 layers}

\begin{figure}
\centering
\includegraphics[width=\columnwidth, keepaspectratio]{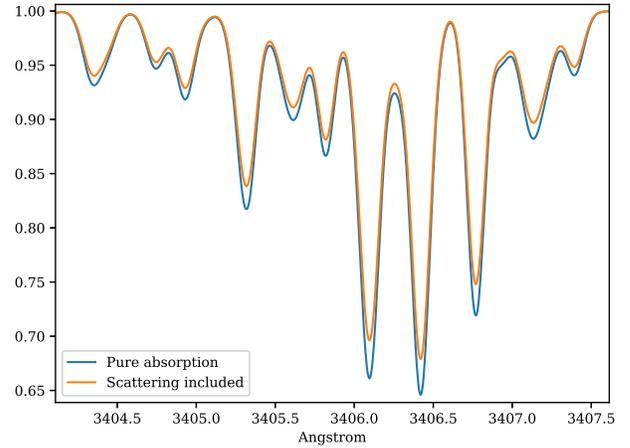}
\caption{Synthetic spectra for star HP1-2115
with ($T_{\rm eff}$=4530, ${\rm log}~g$=2.0, [Fe/H]=-1.0, $v_{\rm t}$=1.45),
illustrating the
change in the continuum, with both absorption and scattering (blue spectrum),
and considering pure absorption (orange spectrum).}
\label{continuum}
\end{figure}

The inclusion of continuum from scattering is taken into account by
adding the scattering opacity from the MARCS models.
This is illustrated in Figure \ref{continuum} for the stellar parameters of star HP1-2115
with ($T_{\rm eff}$=4530, ${\rm log}~g$=2.0, [Fe/H]=-1.0, $v_{\rm t}$=1.45)
 (Barbuy et al. 2016),
in the near-ultraviolet, where this effect is more pronounced.
MARCS opacities are interpolated by \texttt{innewmarcs} using a grid of
atmospheric models with opacities included (\texttt{grid.moo}), generating
file \texttt{modeles.opa}, which is identical in structure to ``.opa'' files
downloaded from MARCS website. A similar calculation was presented in
Figure 1 of Barbuy
et al. (2011), based on calculations by B. Plez using the code Turbospectrum.

%We also compare this Figure with the same calculation carried out with
%Turbospectrum (Figure xx2)

\subsubsection{Atomic line lists}

Atomic line lists can be converted from VALD3 (extended format)
(Piskunov et al. 1995, Ryabchikova et al. 2015) lists of atomic lines.
The VALD atomic line list started with basically the list
by Kur\'ucz (1995, 2017), with progressive implementation of improvements
coming from different sources.
The default atomic line list in the range
3000 - 10.000 {\rm \AA} is from VALD, where the collisional
broadening is obtained by adopting a general formula for the
van der Waals broadening
(e.g. Gray 2005). Better broadening constant values can be obtained through
the code from Anstee \& O’Mara (1995), Barklem
\& O’Mara (1997), Barklem et al. (1998), and Barklem et al.
(2000) (this series of papers is referred to hereafter as ABO)
for neutral lines.

The list of lines in the near-infrared adopted
 is from  Mel\'endez \& Barbuy (1999).
 An updated such line list is given  by Shetrone et al. (2015).

Other options of line lists are available, in particular our own list
that includes damping parameters from ABO for most lines in the
visible region, and
lists of hyperfine structure for several elements, that can be obtained
by contacting the authors.

\subsubsection{Molecular line lists}

The list of molecules, their system, wavelength coverage, number of
vibrational transitions $(v',v'')$, the constants dissociation potential
$D_{0}$ (eV) and the electronic oscillator strength $f_{\rm el}$, that is
related to the transition moment {\bf R}$_{\rm e}$,  with
corresponding references are given in Table \ref{listmol}.
This list of molecular lines is included in the code package,
in file \texttt{molecules.dat}.

In the near-ultraviolet and visible, the line lists from Kur\'ucz
(1993b) 
(http://kurucz.harvard.edu/molecules.html)
were adopted (and transformed to our format) for OH, and CN blue.
CH lines are from Luque \& Crosley (1999).
MgH lines are from Balfour \& Cartwright (1976).
 The NH blue lines are adopted from the recent line lists by Fernando et al.
(2018).
The CN red, and C$_2$ Swan lines  are from 
 C$_2$ laboratory line lists by Phillips \& Davis (1968), and
CN ${\rm A}^{2}\Pi - {\rm X}^{2}\Sigma$ red line lists
by  Davis \& Phillips (1963).
TiO line lists are from Alvarez \& Plez (1998), Plez (1998), 
and Jorgensen (1994) -- see Schiavon et al. (1999).
FeH line lists are from Phillips et al. (1987),
 and constants are described in Schiavon et al. (1997).

 Note that the MgH, CN red,  C$_2$ and FeH adopted correspond
to laboratory measurements, differently to line lists available
in most of the other codes.

In the near-infrared, the CO X$^1 \Sigma$ lines are from Goorvitch (1994).
The OH X$^2 \Pi$ lines were made available by S.P. Davis, that were
originally from Abrams et al. (1994) with a few theoretical lines from
Goldman et al. (1998) -- see Mel\'endez \& Barbuy (1999) and
Mel\'endez et al. (2001).

Next we describe the structure of the PFANT molecular line lists files, which 
are hierarchically organized as follows:

\begin{enumerate}
\item for each molecular system:
constants including the electronic oscillator strength f$_{\rm el}$,
 dissociation constant D$_{0}$ % and mean values of partition function UA, UB

\item for each $(v',v'')$ transition in each molecular system:
      $q_{v'v''}$, $G_0(v)$, $B_{v}$, $D_{v}$
\item for each molecular line in each $(v',v'')$ transition: $\lambda_{0 J'J''}$, $J''$, $(\lskovacs_{\rm norm})_{J'J''}$, where (using Eq. \ref{eq:norm})
\end{enumerate}

\begin{equation}
(\lskovacs_{\rm norm})_{J'J''} = {\lskovacs_{J'J''} \over (2-\delta_{0,\Lambda'+\Lambda''})(2S+1)(2J+1)}
\end{equation}

so that (sum for all branches, fixed $J$)

\begin{equation}
\sum (\lskovacs_{\rm norm})_{J'J''} = 1.
\end{equation}

$G_0(v)$, $B_{v}$, $D_{v}$ are calculated as follows:

\begin{eqnarray*}
B_{v} & = & B_{\rm e} - \alpha_{\rm e} (v+0.5) \\
D_{v} & = & (D_{\rm e} + \beta_{\rm e} (v+0.5)) \times 10^6 \\
G(0) & = & \omega_{\rm e} / 2.0 - \omega_{\rm e}x_{\rm e} / 4.0 +
         \omega_{\rm e}y_{\rm e} / 8.0 \\
G(v) & = & \omega_{\rm e} (v+0.5) - \omega_{\rm e}x_{\rm e}(v+0.5)^2 +
         \omega_{\rm e}y_{\rm e}(v+0.5)^3 \\
G_0(v) & = & G(v) - G(0)
\end{eqnarray*}

\subsubsection{Extending the molecular line list}
\label{webpages}

 Several databases present updated line lists. Some of the most
well-known are described below:

Bertrand Plez's website\footnote{http://www.pages-perso-bertrand-plez.unaiv-montp2.fr}
links to a directory of line lists in various formats.  A constant update
in line lists such as CH (Masseron et al. 2014), and NH (Fernando et al. 2018)
carried out within Turbospectrum,
makes it extremely important to be able to implement these
 new data in other codes
(and our own),
by converting the formats from each other - see below and Sect. 3.3.1.

The Robert L. Kur\'ucz website\footnote{http://http://kurucz.harvard.edu/} compiles several
line lists converted to a standardized format (herein referred to as the ``Kur\'ucz format'').
%\textcolor{red}{\textless--- Kurucz cita muitas referencias, cada readme cita um paper
% diferente, mas nomes recorrentes s~ao Brooke, Bernath, Masseron etc.}

The High Resolution Transmission (HITRAN) brings molecular line lists, 
with description updated in Rothman et al. (2013).

 As concerns complete line lists, the drawback is that they can reach millions of
theoretical lines, with most of them being weak. The situation is similar to the millions
of weak atomic lines (Kur\'ucz 1995). It is not our intent here to cover
all these weak lines. This is accomplished for example by
Tennyson \& Yurchenko (2012), where laboratory and earlier calculations of
molecular lines are completed by ab inition quantum mechanical treatment.
Their line lists are presented in a standardized format using Einstein
coefficients, in the EXOMOL database
\footnote{http://www.exomol.com/data/molecules}.

The webpage by Peter Bernath\footnote{http://bernath.uwaterloo.ca/molecularlists.php}
contains a database on papers regarding a series of molecules.

The webpage by Jeremy Bailey\footnote{http://newt.phys.unsw.edu.au/\~jbailey/vstar\_mol.html} brings line lists, links to other databases,
and software tools.

The {\it Virtual Atomic and Molecular Data Centre Consortium} -
VAMDC\footnote{http://www.vamdc.edu} gives access to 36 inter-connected
atomic and molecular databases (Dubernet et al. 2016).

Given all the different formats of the line lists available,
line list conversion codes are valuable resources,
as they allow for validations of synthesis codes and exchanges 
among research groups.
We implemented a tool to convert molecular line lists from the Kur\'ucz
formats (old and new), and also from the Turbospectrum/BSYN
format, to the PFANT format, as described in Sect. \ref{convmol}.
Other conversion paths are planned.

\begin{table*}
\scalefont{0.85}
\caption{Molecular lines included in PFANT and respective sources of data:
line lists, dissociation potential D$_{o}$, electronic oscillator strength
f$_{\rm el}$.}
\begin{tabular}{ l c r r r c r r }
\hline
\\
Molecule      & System                        & No. lines & $\lambda\lambda$(nm) & No. $(v',v'')$ &
 Linelists & $D_{\circ}$(eV) & $f_{\rm el}$ (ref.)  \\
\\
\hline
\\
MgH           & A$^2$$\Pi - $X$^2$$\Sigma$    & 1,945   & 470$-$609    & 13  & (1)     &1.34 (14)  & 0.2793  (17)  \\
C$_2$ Swan    & d$^3$$\Pi - $a$^3$$\Pi$       & 11,254  & 428$-$677    & 35  & (2)     &6.21 (15)  & 0.033  (18)  \\
CN~blue       & B$^2$$\Sigma - $X$^2$$\Sigma$ & 92,851  & 300$-$600    & 197 & (3)     &7.72 (16)  & 0.0388  (19)  \\
CN~red        & A$^2$$\Pi - $X$^2$$\Sigma$    & 23,828  & 483$-$2715   & 88  & (4)     &"          & 0.00676   (20) \\
CH-AX         & A$^2$$\Delta - $X$^2$$\Pi$    & 10,137+13,830$^\dagger$  & 321$-$787 & 20+31$^\dagger$ & (5) &3.46 (14)  & 0.005257  (21)  \\
CH-BX         & B$^2$$\Sigma - $X$^2$$\Pi$    & 2,016   & 360$-$683    & 10  & (5)     &"          & 0.0025  (22)  \\
CH-CX         & C$^2$$\Sigma - $X$^2$$\Pi$    & 2,829   & 269$-$425    & 12  & (5)     &"          & 0.00595  (23)  \\
CO-NIR        & X$^1$$\Sigma^+$               & 7,088   & 1557$-$5701  & 63  & (6)     &11.09 (14) & --  (24)  \\
NH~blue       & A$^3$$\Pi - $X$^3$$\Sigma$    & 8,599   & 300$-$600    & 55  & (7)     &3.47 (14)  & 0.008 (22) \\
OH~blue       & A$^2$$\Sigma - $X$^2$$\Pi$    & 6,018   & 300$-$540    & 46  & (3)     &4.39 (14)  & 0.0008 (22)  \\
OH-NIR        & X$^2$$\Pi$                    & 2028    & 746$-$25294  & 43  & (8,9)   &"          & --  (24)  \\
FeH           & A$^4$$\Delta - $X$^4$$\Delta$ & 2,705   & 778$-$1634   & 9   & (10)    &1.63 (16)   & 0.001  (25) \\
TiO $\gamma$  & A$^3$$\Phi - $X$^3$$\Delta$   & 26,007  & 622$-$879    & 23  & (11,12,13) &6.87 (14)   & 0.15  (11)  \\
TiO $\gamma'$ & B$^3$$\Pi - $X$^3$$\Delta$    & 219,367 & 500$-$915    & 81  & (11,12,13) &"           & 0.14  (11)  \\
TiO $\alpha$  & C$^3$$\Delta - $X$^3$$\Delta$ & 360,726 & 388$-$837    & 79  & (11,12,13) &"           & 0.12  (11)  \\
TiO $\beta$   & c$^1$$\Phi - $a$^1$$\Pi$      & 91,805  & 431$-$796    & 63  & (11,12,13) &"           & 0.006  (11)  \\
TiO $\delta$  & b$^1$$\Pi - $a$^1$$\Delta$    & 189,019 & 622$-$1480   & 66  & (11,12,13) &"           & 0.05  (11)  \\
TiO $\epsilon$& E$^3$$\Pi - $X$^3$$\Delta$    & 253,755 & 641$-$1341   & 61  & (11,12,13) &"           & 0.014  (11)  \\
TiO $\phi$    & b$^1$$\Pi - $d$^1$$\Sigma$    & 105,082 & 664$-$1800   & 65  & (11,12,13) &"           & 0.052  (11)  \\
\hline
\label{listmol}
\end{tabular}
\textsuperscript{}
\\
\footnotesize{References:
(1) Balfour \& Cartwright (1976);
 (2) Phillips \& Davis (1968); (3) Kur\'ucz (1993);
(4) Davis \& Phillips (1963);  (5) Luque \& Crosley (1999);
 (6) Goorvitch (1994); (7) Fernando et al. (2018); (8) Abrams et al. (1994);
(9) Goldman et al. (1998); (10) Phillips et al. (1987);
(11) Alvarez \& Plez (1998); (12) Jorgensen (1994); (13) Plez (1998);
 (14) Huber \& Herzberg (1979);
(15) Pradhan et al. (1994);
(16) Schulz \& Armentrout (1991); (17) Henneker \& Popkie (1971);
(18) Kirby et al. (1979); (19) Duric et al. (1978); (20) Bauschlicher et al. (1988);
(21) Brzozowski et al. (1976); (22) Grevesse \& Sauval (1973); (23) Lambert (1978);
(24) Goldman et al. (1998); (25) Schiavon et al. (1997). \\
$^\dagger$12CH + 13CH. \\
 }
 \end{table*}

\subsubsection{Output spectra}

The code
\texttt{pfant} generates three output files corresponding to the
continuum, un-normalized flux, and normalized flux. As usual, the normalized flux is
$flux_{\rm norm}(\lambda) = flux_{\rm un{\text -}normalized}(\lambda) / flux_{\rm continuum}(\lambda)$.

The code
\texttt{nulbad} takes \texttt{pfant} output spectra, convolves it with
a Gaussian profile (GP), and generates a file that, by default, contains
the FWHM (full width at half maximum, given in \AA) of the Gaussian profile
in its name. This is a two-column text file with the wavelength (\AA) and
flux.

\subsection{Python interface and tools}
\label{secpyfant}

With package \texttt{pyfant} (http://trevisanj.github.io/pyfant),
one can write scripts that use PFANT spectral synthesis in the core
of more complex tasks
(e.g. automatic determination of stellar parameters)
in the Python language.
The \texttt{pyfant} package API (application programming interface)
includes resources to run PFANT spectral synthesis in series and in parallel,
and to load, filter, transform, save and visualize relevant data files.

In addition, \texttt{pyfant} contains a series of command-line and graphical tools
(command-line and graphical interfaces) to perform
various tasks, including a scraper of molecular constants from the NIST
chemistry web book, an atomic lines converter from VALD3 to PFANT format, a
molecular line list converter to the PFANT format,
a tool to create grids of MARCS models, an editor for atomic lines,
and an editor for molecular lines. A complete list of tools is obtained
by running script \texttt{programs.py} in the computer terminal.

\subsubsection{Conversion of molecular line lists} \label{convmol}

A systematic method to convert molecular line lists to the PFANT format
was implemented. Conversion can be carried out with very little user
interaction, provided that:

\begin{itemize}
  \item Diatomic molecular constants are available in the NIST chemistry
        web book for the particular molecular systems of choice\footnote{NIST is a
        convenient source because their data can be automatically retrieved, using the
         tool \texttt{moldbed.py}. However in some cases the NIST database does not contain the most recent data,
        so it is of interest to consult other databases, such as reported in Sect. \ref{webpages}.}
  \item The input line list format contains sufficient information to
        determine the branch of each line.
\end{itemize}

 In this section we discuss some particularities with the conversions of the
Kur\'ucz and Turbospectrum line lists. For the Kur\'ucz format,
the branch can be determined from the tabulated $J'$, $J''$, $spin'$, and $spin''$
as follows:

\begin{itemize}
  \item ``P'' if $J'' > J'$;
  \item ``Q'' if $J'' = J'$;
  \item ``R'' if $J'' < J'$.
\end{itemize}

And in more detail if it is ``P1'', ``P2'', ``P3'', etc. is indicated
by the spin of the lower state and upper states.
If $spin'' = spin'$, the branch is ``(P/Q/R)($spin$)'',
If $spin''\neq spin$, the branch is ``(P/Q/R)($spin'$)($spin''$)''.

 Turbospectrum linelists already contain the branch given as a string, e.g. ``R23''.

The branch is used to determine the specific formula from Kov\'acs (1969) that
will be used to calculate $\lskovacs_{J'J''}$. This choice of formula is a detailed
procedure, and depends on:

\begin{itemize}
	\item the multiplicity of the transition (singlet, doublet, triplet);
	\item the sign and value of $\Delta\Lambda = \Lambda' - \Lambda''$;
	\item the sign of $A$; and
	\item the branch.
\end{itemize}

The Franck-Condon factors $q_{v'v''}$ (eq. 16) are computed with
the code for transition probabilities of molecular transitions
by Jarmain \& McCallum (1970) 
(TRAPRB).%\footnote{http://github.com/trevisanj/traprb}. 
The $q_{v'v''}$ factors
are dependent on the system, and the vibrational levels, and
have almost no dependence on rotational constants
(Singh \& de Almeida 1982).

The procedure described above  to convert molecular linelists is coded as part of the \texttt{pyfant} API,
and is available as a graphical application named \texttt{convmol.py}. This
application interfaces with the NIST chemistry web book to get the diatomic
molecular constants $A$, $B_{\rm e}$, $\alpha_{\rm e}$, $\beta_{\rm e}$, $\omega_{\rm e}$,
$\omega_{\rm e}x_{\rm e}$, and $\omega_{\rm e}y_{\rm e}$ (these can be later tuned
if necessary), and interfaces with code TRAPRB to calculate $q_{v'v''}$.

\subsection{Comparison with stellar spectra and code Turbospectrum}

% A validation of the code PFANT is presented here, with the comparison
% the  CN blue ${\rm B}^2 \Sigma - {\rm X}^2 \Sigma$ ($v'=0$,$v''=0$)
% bandhead, and the
% CH ${\rm A}^2 \Delta - {\rm X}^2 \Pi$ (0,0) G-bandhead, of the spectra resulting from
% the calculations with PFANT, Turbospectrum, and the
% observed solar spectrum (Kur\'ucz et al. 1984).

\begin{figure}
\centering
\includegraphics[width=\columnwidth, keepaspectratio]{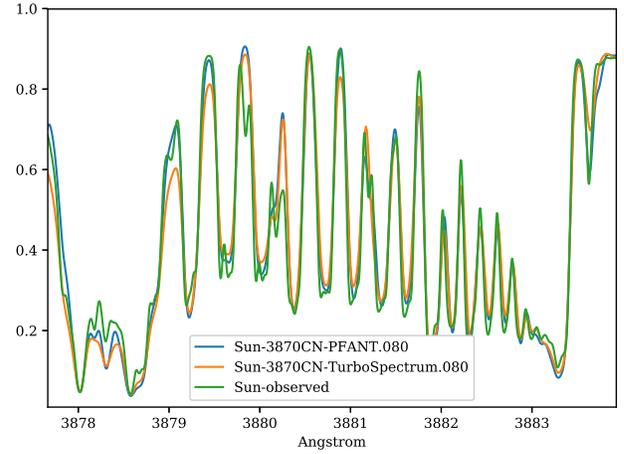}
\caption{Synthetic and observed spectra of the the Sun,
in the region of the CN  ${\rm B}^2\Sigma - {\rm X}^2\Sigma$ (0,0)
bandhead: observed spectrum (green);
synthetic spectra with Turbospectrum (orange), and with PFANT (blue).}
\label{suncn}
\end{figure}

\begin{figure}
\centering
\includegraphics[width=\columnwidth, keepaspectratio]{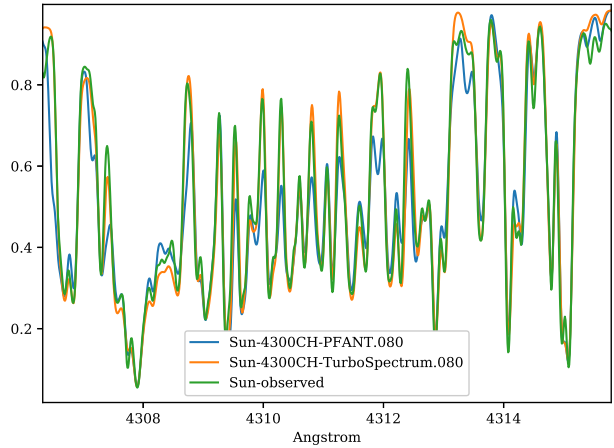}
\caption{Synthetic and observed spectra of the the Sun,
in the region of the CH A$^2\Delta - $X$^2\Pi$ (0,0) G-bandhead: observed spectrum (green);
synthetic spectra with Turbospectrum (orange), and with PFANT (blue).}
\label{sunch}
\end{figure}

\begin{figure}
\centering
\includegraphics[width=\columnwidth, keepaspectratio]{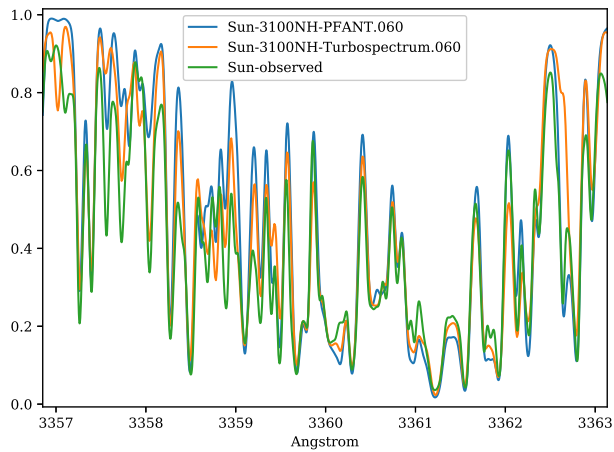}
\caption{Synthetic and observed spectra of the the Sun,
in the region of the NH A$^3\Pi - $X$^3\Sigma$ (0,0) bandhead: observed spectrum (green);
synthetic spectra with Turbospectrum (orange), and with PFANT (blue).}
\label{sunnh}
\end{figure}

A validation of the code PFANT is presented here, with the comparison
of the
CN blue ${\rm B}^2 \Sigma - {\rm X}^2 \Sigma$ ($v'=0$,$v''=0$) bandhead (Figure \ref{suncn}), the
CH ${\rm A}^2 \Delta - {\rm X}^2 \Pi$ (0,0) G-bandhead (Figure \ref{sunch}), and the
NH A$^3\Pi - $X$^3\Sigma$ (0,0) bandhead (Figure \ref{sunnh}) 
of the spectra resulting from
the calculations with PFANT, Turbospectrum, and the
observed solar spectrum (Kur\'ucz et al. 1984).

\section{Summary}
\label{secsummary}

 We present the description of calculation of diatomic molecular lines,
given that this task is not straightforward. 
Molecular line intensities in the literature are given
 in either of the two formats:
 the Einstein $A$ constant, related to the oscillator strength 
(see Sect. 2), or else the inclusion of all the ingredients that 
build together the oscillator strength is needed.

We present a description of a new  updated online version
of the code for spectrum synthesis PFANT, together with
many accessory tools and documentation. We describe the
molecular line lists adopted, recalling that
they need constant updating and implementation of new
molecules. One example of an important update is the 
recent line list for NH (Fernando et al. 2018).
The package includes codes for conversion of molecular line lists to the PFANT format,
in particular so far from Kur\'ucz's and Plez's line lists.
%Further line lists to be implemented would be CaH, AlH.
The atomic line list is also in constant improvement, both from the VALD3 database,
as our own line list (available upon request).

\begin{acknowledgements}
We are grateful to the referee for constructive comments that
helped improving the code and the paper.
We thank J. Mel\'endez and R. Smiljanic for useful comments
on a previous version of the manuscript.
BB and AA acknowledge grants from CAPES - Finance code 001,
 CNPq and FAPESP.
JT acknowledges the TT-5 FAPESP fellowships 2014/25987-5
and 2016/00777-3.
\end{acknowledgements}

%\begin{appendix}

%\section{AN EXAMPLE OF APPENDIX HEAD}
%xxxx

\label{lastpage}

\begin{thebibliography}{}
\bibitem[]{} Abrams, M. C., Davis, S. P., Rao, M. L. P., Engleman, R., Jr.,
 Brault, J.W. 1994, ApJS, 93, 351
\bibitem[]{} Alvarez, R., Plez, B. 1998, A\&A, 330, 1109
\bibitem[]{} Anstee, S.D., O'Mara, B.J. 1995, MNRAS, 276, 859
\bibitem[Asplund et al.(2009)]{asplund09}  Asplund, M., Grevesse, N.,
 Sauval, A.J., Scott, P. 2009, ARA\&A, 47, 481
\bibitem[]{} Barbuy, B. 1982, PhD Thesis, Universit\'e de Paris VII
\bibitem[]{} Barbuy, B., Perrin, M.-N., Katz, D., Coelho, P.,
 Cayrel, R., Spite, M.,  Van't Veer-Menneret, C. 2003, A\&A,  404, 661
\bibitem[]{} Barbuy, B., Spite, M., Hill, V. et al. 2011, A\&A, 534, A60
\bibitem[]{} Barbuy, B., Cantelli, E., Vemado, A. et al. 2016, A\&A, 591, A53
\bibitem[]{} Barklem, P.S., O'Mara, B.J. 1997, MNRAS, 290, 102
\bibitem[]{} Barklem, P.S., Anstee, S.D., O'Mara, B.J. 1998, PASA, 15, 336
\bibitem[]{} Barklem, P.S., Piskunov, N.E., O'Mara, B. J. 2000, A\&AS, 142, 467
\bibitem[]{} Balfour, W.J., Cartwright, H.M. 1976, A\&AS, 26, 389
\bibitem[]{} Bauschlicher, C. W., Langhoff, S. R., Taylor, P. R. 1988, ApJ, 332, 531
\bibitem[]{} Biehl, D. 1976, PhD Thesis, Kiel University
\bibitem[\protect\citeauthoryear{Born \& Oppenheimer}{1927}]{1927AnP...389..457B} Born M., Oppenheimer R., 1927, AnP, 389, 457
\bibitem[]{} Brzozowski, J., Bunker, P., Elander, N., Erman, P. 1976, ApJ, 207, 414
\bibitem[]{} Castilho, B.V., Spite, F., Barbuy, B., Spite, M., de Medeiros, J.R., 
  Gregorio-Hetem, J. 1999, A\&A, 345, 249
%\bibitem[]{} Cayrel, R. 1960, Ann. Ap. 23, 233
\bibitem[]{} Cayrel, R., Perrin, M.-N., Barbuy, B., Buser, R. 1991, A\&A, 247, 108
\bibitem[]{} Coelho, P., Barbuy, B., Mel\'endez, J., Schiavon, R.P.,
 Castilho, B.V.  2005, A\&A, 443, 735
\bibitem[]{} Davis, S.P., Phillips, J.G. 1963, The Red System
(A$^{2}\Pi$-X$^{2}\Sigma$) of the CN molecule
\bibitem[]{} Dubernet, M.-L., Zw\"olf, C.M., Moreau, N., Ba, Y.A. 2016, IAUFM, 29A, 347
\bibitem[]{} Duric, N., Erman, P., Larsson, M. 1978, Phys. Scr., 18, 39
\bibitem[]{} Edvardsson, B., Andersen, J., Gustafsson, B.,
Lambert, D.~L., Nissen, P.~E., 
  Tomkin, J. 1993, A\&A, 275, 101
\bibitem[\protect\citeauthoryear{Einstein}{1917}]{1917PhyZ...18..121E} Einstein A., 1917, PhyZ, 18, 121
%\bibitem[]{} Einstein, A. 1917, PhyZ, 18, 121
\bibitem[]{} Fernando, A.M., Bernath, P.F., Hodges, J.N., Masseron, T. 2018,
JQSRT, 217, 29
\bibitem[]{} Gray, D.F. 2005, The Observation and Analysis of Stellar
Atmospheres, 3rd edition, Cambridge U. Press, UK
\bibitem[]{} Grevesse, N., Noels, A., Sauval, A.J. 1996, Standard
Abundances, in {\it Cosmic abundances}, editors: Holt, S.~S. and 
Sonneborn, G., Astronomical Society of the Pacific Conference Series,
99, 117
\bibitem[]{} Grevesse N., Sauval J., 1973, A\&A 27, 29
%\bibitem[Grevesse \& Sauval(1998)]{grevesse98}
%Grevesse, N. \& Sauval, J.N. 1998, SSRev, 35, 161
\bibitem[]{} Grevesse, N., Asplund, M., Scott, P., Sauval, A.J.
2015, A\&A, 573, A27
\bibitem[]{} Goldman, A., Schoenfeld, W.~G., Goorvitch, D., Chackerian, Jr., C.,
   Dothe, H., M{\'e}len, F., Abrams, M.~C., Selby, J.~E.~A. 1998, JQSRT, 59, 453
\bibitem[]{} Goorvitch, D. 1994, ApJS, 95, 535
\bibitem[]{} Gustafsson, B., Edvardsson, B., Eriksson, K., 
 J\"orgensen,  U.~G.,   Nordlund, {\AA}., Plez, B. 2008, A\&A, 486, 951
%\bibitem[]{} Heiter, U., Lind, K., Asplund, M. et al. 2015,
%Physica Scripta, 90, Issue 5 (arXiv:1506.06697)
\bibitem[]{} Henneker, W.H. Popkie, H.E. 1971, J. Chem. Phys. 54, 1763
\bibitem[]{} H\"onl, H.,  London, F. 1925, Zs. f. Physik, 33, 803
\bibitem[]{} Huber, K.P., Herzberg, G. 1979,
 {\it Constants of Diatomic Molecules},
      van Nostrand Reinhold Cy., New York.
\bibitem[]{} Hougen, J.T. 1970, Nat. Bur. Stand. Monog. 115
\bibitem[]{} Jarmain, W.R., McCallum, J.C. 1970, 
 Technical Report: TRAPRB, A Computer Programme for Molecular Transitions, 
Department of Physics, University of Western Ontario, December 1970.
\bibitem[]{} Jorgensen, U.G. 1994, A\&A, 284, 179
\bibitem[]{} Kirby, K., Saxon, R. P., Liu, B. 1979, ApJ, 231, 637
\bibitem[]{} Kov\'acs, I. 1969, {\it Rotational Structure in the Spectra of
    Diatomic Molecules}, Adam Hilger Ltd. London
\bibitem[]{} Kur\'ucz R. L. 1970, SAO Special Report, 309
\bibitem[]{} Kur\'ucz R. L., Furenlid I., Brault J., Testerman L.,
 1984, National Solar
Observatory Atlas, National Solar Observatory, Sunspot, NM
\bibitem[]{} Kur\'ucz, R.L. 1993a, CD-ROM No. 18. Cambridge, Mass.: Smithsonian Astrophysical Observatory
\bibitem[]{} Kur\'ucz, R.L. 1993b, CD-ROM No. 15. Cambridge, Mass.: Smithsonian Astrophysical Observatory
\bibitem[]{} Kur\'ucz, R.L.
1995, Atomic Line Data (R.L. Kurucz and B. Bell)
Kurucz CD-ROM No. 23. Cambridge, Mass.: Smithsonian Astrophysical Observatory
\bibitem[]{} Kur\'ucz, R.L.
1993, Diatomic Molecular Data for Opacity Calculations. Kurucz
CD-ROM No. 15. Cambridge, Mass.: Smithsonian Astrophysical Observatory
\bibitem[]{} Kur\'ucz, R.L. 2017, Can. J. Phys. 95, 825
\bibitem[]{} Lambert D.L., 1978, MNRAS 182, 249
%\bibitem[]{} Langhoff, S.R., Bauschlicher, C.W., Jr. 1990, J. Mol. Spec. 141, 243
\bibitem[]{} Luque, J., Crosley, D.R. 1999, SRI report no. MP 99-099
%\bibitem[]{} Martin, W.C., Fuhr, J.R., Kelleher, D.E., et al. 2002,
%        NIST Atomic Database (version 2.0),
%        National Institute of Standards and Technology, Gaithersburg, MD.
%\bibitem[]{} Masseron, T. Proceedings of the Annual meeting of the
% French Society of Astronomy and Astrophysics. Eds.: F. Martins, S. Boissier, V%. Buat, L. Cambr\'esy, P. Petit, pp.303-305
\bibitem[]{} Masseron, T., Plez, B., Van Eck, S.,
et al. 2014, A\&A, 571, A47
% Colin R, Daoutidis I, Godefroid M, Coheur P F, Bernath P, Jorissen
\bibitem[]{} Mel\'endez, J., Barbuy, B. 1999, ApJS, 124, 527
\bibitem[]{} Mel\'endez, J., Barbuy, B., Spite, F. 2001, ApJ, 556, 858
%\bibitem[]{} Mel\'endez, J., Barbuy, B., 2009, A\&A, 497, 611
\bibitem[]{} Pasquini, L., Ecuvillon, A., Bonifacio, P., Wolff, B.
      2008, A\&A, 489, 315
\bibitem[]{} Phillips, J.G., Davis, S.P. 1968, The Swan system of
the C$_2$ molecule (U. of California Press)
\bibitem[]{} Phillips, J.G., Davis, S.P., Lindgren, B., Balfour, W.J.
1987, ApJS, 65, 721
\bibitem[]{} Piotto, G., et al. 2015, AJ, 149, 91
\bibitem[Piskunov et al.(1995)]{pisk95}
Piskunov, N., Kupka, F., Ryabchikova, T., T.~A., Weiss, W.~W.,Jeffery, C.~S.
 1995, A\&AS, 112, 525
\bibitem[]{} Piskunov, N.,  Valenti, J.A. 2017, A\&A, 597, A16
\bibitem[]{} Plez, B. 1998, A\&A, 337, 495
\bibitem[]{} Popovas, A. 2014, Master thesis, Niels Bohr Institute
\bibitem[]{} Praderie, F. 1967, Ann. Ap. 30, 31

\bibitem[\protect\citeauthoryear{Pradhan, Partridge, \& Bauschlicher}{1994}]{1994JChPh.101.3857P} Pradhan A.~D., Partridge H., Bauschlicher C.~W., Jr., 1994, JChPh, 101, 3857
\bibitem[]{} Rothman, L.S., Gordon, I.E., Babikov, Y. et al. 2013, JQSRT, 130, 4
\bibitem[\protect\citeauthoryear{Ryabchikova et al.}{2015}]{2015PhyS...90e4005R} Ryabchikova T., Piskunov N., Kurucz R.~L., Stempels H.~C., Heiter U., Pakhomov Y., Barklem P.~S., 2015, PhyS, 90, 054005
\bibitem[]{} Schadee, A. 1964, BAN, 17, 311
\bibitem[]{} Schadee, A. 1967, JQSRT, 7, 169
\bibitem[]{} Schadee, A. 1971, A\&A, 14, 401
\bibitem[]{} Schadee, A. 1975, A\&A, 41, 203
\bibitem[]{} Schadee, A. 1978, JQSRT, 19, 451

\bibitem[\protect\citeauthoryear{Schiavon, Barbuy, \& Singh}{1997}]{1997ApJ...484..499S} Schiavon R.~P., Barbuy B., Singh P.~D., 1997, ApJ, 484, 499
\bibitem[\protect\citeauthoryear{Schiavon \& Barbuy}{1999}]{1999ApJ...510..934S} Schiavon R.~P., Barbuy B., 1999, ApJ, 510, 934

\bibitem[\protect\citeauthoryear{Schultz \& Armentrout}{1991}]{1991JChPh..94.2262S} Schultz R.~H., Armentrout P.~B., 1991, JChPh, 94, 2262
\bibitem[]{} Scott, P.,  Grevesse, N., Asplund, M., et al. 2015a, A\&A, 573, A25
\bibitem[]{} Scott, P., Asplund, M., Grevesse, N., Bergemann, M.
  Sauval, A.J. 2015b, A\&A, 573, A26
\bibitem[]{2015ApJS..221...24S} Shetrone M., et al., 2015, ApJS, 221, 24 
\bibitem[]{} Singh, P.D., de Almeida, A. 1982, JQSRT, 27, 471
\bibitem[]{} Smiljanic, R., Gauderon, R., North, P., Barbuy, B., 
Charbonnel, C.,    Mowlavi, N. 2009, A\&A, 502, 267
\bibitem[]{} Smith, V.V., Terndrup, D.M., Suntzeff, N.B. 2002, ApJ 579, 832
\bibitem[]{} Spite, M. 1967, An. Ap. 30, 211
\bibitem[]{} Spite, M., Cayrel, R., Plez, B. et al. 2005, A\&A, 430, 655

\bibitem[]{} Sneden, C. 1973, PhD thesis, U. of Texas at Austin
\bibitem[]{} Steffen, M., Prakapavicius, D., Caffau, E., Ludwig, H.-G., 
Bonifacio, P., Cayrel, R., Kučinskas, A., Livingston, W. C.  2015,
A\&A, 583, A57
\bibitem[]{} Tatum, J.B. 1967, ApJS, 14, 21
\bibitem[]{} Tennyson, J., Yurchenko, S.N. 2012, MNRAS, 425, 21
\bibitem[]{} Tsuji, T. 1973, A\&A, 23, 411
\bibitem[]{} Trevisan, M., Barbuy, B., Eriksson, K.,
Gustafsson, B., Grenon, M., Pomp\'eia, L.
2011, A\&A, 535, A42
\bibitem[]{} Valenti, J.A., Piskunov, N. 1996, A\&AS, 118, 595
\bibitem[]{} Whiting E.E., Nicholls, R.W. 1974, ApJS, 27, 1
\bibitem[]{} Whiting E.E., Schadee, A., Tatum, J.B., Hougen, J. T., 
Nicholls, R. W. 1980, J. Mol. Spec., 80, 249

\end{thebibliography}
\end{document}